\documentclass[a4paper]{jpconf}
\usepackage{graphicx}
\def\be{\begin{equation}}
\def\ee{\end{equation}}

\def\bea{\begin{eqnarray}}
\def\eea{\end{eqnarray}}
\newcommand{\ket}[1]{|{#1}\rangle}

\begin{document}
\title{Rydberg state mediated quantum gates and entanglement of pairs of neutral atoms}

\author{M Saffman, X L Zhang,  A T Gill, L Isenhower, and T G Walker}

\address{Department of Physics, University of Wisconsin, 1150 University Avenue, Madison, Wisconsin 53706 USA}

\ead{msaffman@wisc.edu}

\begin{abstract}Experiments performed within the last year have demonstrated Rydberg state mediated quantum gates and deterministic entanglement between pairs of trapped neutral atoms. These experiments validate ten year old proposals for Rydberg mediated quantum logic, but are only the beginning of ongoing efforts to improve the fidelity of the results obtained and scale the experiments to larger numbers of qubits. We present here a summary of the results to date, along with a critical evaluation of the prospects for higher fidelity Rydberg gates. 
\end{abstract}

\section{Introduction}

One of the central challenges in building devices suitable for quantum logic operations is the need to isolate  quantum information from environmental decoherence while simultaneously allowing for strong and controllable interactions between pairs of qubits. Neutral atoms in optical traps formed by far off-resonance laser beams represent one of the best isolated systems known. Quantum information can be stored in long-lived hyperfine states, with decoherence times that have been measured to be as long as several seconds\cite{Treutlein2004}. The challenge for neutral atom based quantum computing has centered around the difficulty of introducing a strong and controllable two-atom interaction. A range of ideas have been proposed \cite{Brennen1999,Jaksch1999,Pellizzari1995,You2000,Lukin2000,Mompart2003}
and collisional entanglement\cite{Jaksch1999} has been demonstrated in many atom optical lattice experiments\cite{Mandel2003,Anderlini2007}. 

In 2000 Jaksch, et al.\cite{Jaksch2000} introduced the idea of using strong dipolar interactions between Rydberg excited atoms for  
quantum gates. If a ``control" atom is Rydberg excited with a resonant laser pulse, subsequent excitation of a second proximal ``target" atom will be off-resonant, and therefore blocked, due to dipolar interactions between the Rydberg atoms. This Rydberg blockade form of gate operation provides a straightforward way of implementing a controlled $\pi$ phase shift on the target atom using a three pulse sequence (read from right to left): $U_Z=  U_{r\rightarrow g}^{(c)}(\pi)
U_{g\rightarrow r\rightarrow g}^{(t)}(2\pi)U_{g\rightarrow r}^{(c)}(\pi). $
Here $U^{(c/t)}(\theta)$ is a Rabi pulse of area $\theta$ applied between ground $(|g\rangle)$ and Rydberg $(|r\rangle)$
states of the control (c) or target (t) atom. This sequence gives a $U_Z$ controlled phase gate, which can be readily converted into a controlled-NOT (CNOT) gate by adding single qubit operations on the 
target atom before and after the pulse sequence\cite{Nielsen2000}.
We note that Rydberg states are also suitable for engineering entanglement between non-trapped atoms as was demonstrated in 1997 in cavity QED experiments\cite{Hagley1997}.

The Rydberg blockade gate has a number of desirable characteristics. 
As long as the dipolar blockade interaction ${\sf B}$ is very large compared to the excitation Rabi frequency $\Omega$ the gate works well. 
This is true even if the exact value of $\sf B$ is not known. This means that the separation of the atoms does not have to be controlled very precisely which adds to the robustness of the gate. Furthermore
when ${\sf B}\gg \Omega$ the target atom is never excited to a Rydberg state during the gate. This implies that the strong dipolar interaction does not lead to any energy shifts or forces between the atoms during the gate operation. Remarkably we can have a very strong two-atom interaction without the atoms involved ever experiencing any forces.  

In addition to these desirable features what really sets the Rydberg gate apart from other approaches is the strength of the interaction. 
Consider a pair of alkali atoms at $R=10~\mu\rm m$ separation. The interaction energy for ground state atoms at this length scale is dominated by the magnetostatic dipole-dipole interaction with a strength
$U_{gg}/h\sim 10^{-5}~\rm Hz$. The same atom pair excited to $n=100$ Rydberg states has $U_{rr}\sim 10^7~\rm Hz$, an increase by 12 orders of magnitude\cite{Saffman2010}! The ability to coherently turn the interaction on and off with a contrast of 12 orders of magnitude is possibly unique to the Rydberg system, and is a superb figure of merit for Rydberg mediated quantum logic. The intrinsic properties of the Rydberg blockade approach suggest that entangling gates should be possible with errors at the $10^{-3}$ level or better\cite{Saffman2005a,Saffman2010}.

Initial observations of Rydberg mediated excitation suppression were made in 2004\cite{Singer2004,Tong2004}. Those experiments were done in many atom samples and it was not possible to verify that excitation was limited to a single atom. Rydberg blockade at the level of single atoms was demonstrated by groups from the University of Wisconsin and Institut d'Optique in 2009\cite{Urban2009,Gaetan2009}. In addition to blockade the Institut d'Optique experiment showed that the effective Rabi frequency when simultaneously addressing two atoms was increased by 
a factor of $\sqrt2$, compared to the single atom rate. The reason for this is as follows. In an $N$ atom ensemble subject to strong blockade  an effective two-level system is formed between the states 
$|g\rangle=|g_1g_2...g_N\rangle$ and $|{\sf s}\rangle = \frac{1}{\sqrt{N}}\sum_{=1,N}|g...r_j...g\rangle$. The state $|\sf s\rangle$ has one unit of excitation symmetrically shared between all $N$ atoms. This 
entangled state is coupled to $|g\rangle$ with the Rabi frequency  $\Omega_N=\sqrt{N}\Omega$ which is $\sqrt{N}$ larger than the one-atom coupling\cite{Lukin2001}.  Observation of a $\sqrt2$ speed up by Ga\"etan, et al.\cite{Gaetan2009} implied preparation of the entangled state $|\psi\rangle=\frac{1}{\sqrt2}(|gr\rangle+e^{\imath\varphi'}|rg\rangle))$ and  was thus the first tantalizing observation of Rydberg blockade mediated entanglement. However, the entanglement in this case involves short lived Rydberg states which are not well suited for scalable quantum logic. 

The next step was to map the Rydberg entanglement onto long lived ground states with an additional laser pulse, $|r\rangle\rightarrow|g_1\rangle$. One thereby creates the state 
\begin{equation}
|\psi\rangle=\frac{1}{\sqrt2}(|01\rangle+e^{\imath\varphi}|10\rangle)
\label{eq.Bell}
\end{equation}
 where $0,1$ correspond to  different long lived hyperfine ground states of $^{87}$Rb ($F=1,2$) . This was done by Wilk, et al. \cite{Wilk2010} in 2010. The  fidelity of the expected entangled Bell state was extracted from measurements of parity oscillations\cite{Gaetan2010} and was found to be under the entanglement threshold of $F=0.5$. Correction of the data for loss of atoms during the preparation sequence revealed a fidelity of $F=0.74$  
for the 62\% of the trials that had two atoms remaining after the preparation sequence. Note that the phase $\varphi$ appearing in Eq. (\ref{eq.Bell}) may be  stochastic due to several experimental issues. This limits the Bell state fidelity, a point which we will return to below. 

Independently of the Institut d'Optique experiments the Wisconsin group 
implemented the $U_Z$ operation between two trapped atoms and added $\pi/2$ pulses on the target atom to generate a full CNOT gate matrix.
The CNOT gate was also demonstrated using a different protocol referred to as a controlled amplitude swap\cite{Isenhower2010}.
 The 
CNOT gate was then used with the control atom in the superposition 
state $\frac{1}{\sqrt2}(|0\rangle+i|1\rangle)$ and the target atom in either of the two hyperfine states to create the Bell states 
$|B_1\rangle=\frac{1}{\sqrt2}(|00\rangle+|11\rangle)$ or 
$|B_2\rangle=\frac{1}{\sqrt2}(|01\rangle+|10\rangle)$\cite{Isenhower2010}.
The fidelity of the Bell state $\ket{B_1}$ was evaluated with data extracted from 
parity oscillation measurements and was under $F=0.5.$ Correcting for the fraction of atom pairs remaining after the preparation steps revealed post-selected entanglement with fidelity $F=0.58$. 

These data were subsequently improved upon using an upgraded apparatus which resulted in Bell state fidelity of $F=0.58\pm0.04$ without applying any corrections for atom loss or trace loss\cite{Zhang2010}, and a loss corrected fidelity of $F=0.71.$ These latest results demonstrate that Rydberg blockade can be used for deterministic entanglement of neutral atom pairs  which is a prerequisite for larger scale quantum information processing. 

The results obtained in the past year are a significant step forwards 
in neutral atom quantum logic. Nevertheless it is also apparent that a large gap exists between what has been achieved and what should be possible using Rydberg blockade. In Sec. \ref{sec.errors}  of this contribution we will take a closer look at the technical and intrinsic errors associated with Rydberg mediated entanglement. The results of this analysis suggest that it should indeed be possible to reach $10^{-3}$ gate errors, or better, but that doing so will not be particularly simple and will require a carefully designed experimental apparatus. 
In Sec. \ref{sec.outlook} we conclude with some remarks on the potential for scaling to a many qubit experiment.

\section{Entangling gate errors}
\label{sec.errors}

Imperfections in quantum gates arise from two categories of errors. Those we term {\it intrinsic} errors  are due to the choice of gate 
protocol and the characteristics of the physical system used to encode the qubits. In addition there are {\it technical} errors which arise from imperfections and noise in the applied pulses, fluctuations of background magnetic or electric fields, and motional effects at finite atomic temperature. A rigorous treatment would be based on including all of these error sources in the Bloch equations, solving them for the $C_Z$ pulse sequence, and then averaging the results over the computational input states. In the limit where all errors are small, which is the situation we are most interested in, a simpler and more physically transparent approach is based on separately estimating the various error sources and then adding them together. We follow this latter approach here.   

Furthermore we will treat only the errors associated with the basic two-qubit entangling operation. A quantum computer requires additional capabilities including state preparation, single qubit operations, and measurements. A full discussion of those errors for a neutral atom implementation is outside the scope of this paper.

\subsection{Intrinsic errors}

The intrinsic error of a Rydberg blockade controlled phase gate can be exceedingly low. When the hyperfine qubit states have a frequency separation $\omega_{\rm hf}$ that is large compared to both $\Omega$ and $\sf B$ the optimum excitation Rabi frequency is\cite{Saffman2010}
$\Omega_{\rm opt}=(7\pi)^{1/3}\frac{{\sf B}^{2/3}}{\tau^{1/3}}$ and the corresponding minimum gate error is $E_{\rm min}=\frac{3(7\pi)^{2/3}}{8}\frac{1}{{({\sf B}\tau})^{2/3}}$, with $\tau$ the Rydberg radiative lifetime at room temperature. Very low gate errors are obtained when both $\sf B$ and $\Omega$ become large and are no longer completely negligible compared to $\omega_{\rm hf}.$ In this case we must use the full expression\cite{Saffman2010}
\be
E\simeq\frac{7\pi}{4\Omega\tau}\left(1+\frac{\Omega^2}{\omega_{\rm hf}^2} +\frac{\Omega^2}{7 {\sf B}^2}\right)
+\frac{\Omega^2}{8{\sf B}^2}\left(1+6 \frac{{\sf B}^2}{\omega_{\rm hf}^2}\right).
\label{eq.Ebar}
\ee

\begin{figure}[!t]
\vspace{-.1cm}
\begin{center}
\includegraphics[width=15.cm]{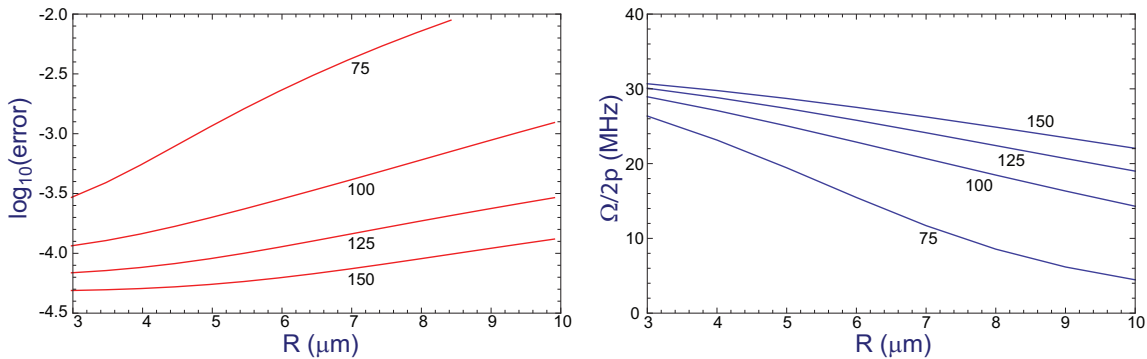}
\end{center}
\vspace{-.7cm}
\caption{\small \label{fig.Cphaseerror} Intrinsic error of the $C_Z$ gate (left) and corresponding Rabi excitation frequency (right)
for Rb $ns_{1/2}$ states with $n=75,100,125,150$. Calculations used radiative lifetimes of $\tau = 180, 340, 570, 860~\mu\rm s$.  }
\end{figure}

Figure  \ref{fig.Cphaseerror} shows $E$ evaluated from Eq. (\ref{eq.Ebar}) at  $\Omega=\Omega_{\rm opt}$ as a function of atom separation $R$ for $^{87}$Rb atoms excited to 
$ns_{1/2}$ states. The blockade shift ${\sf B}(R)$ was calculated using the methods described in \cite{Walker2008}.  We see that the intrinsic error can fall under $10^{-4}$ for $n>100$ although this requires rather high Rabi excitation frequencies. For example the $150s_{1/2}$ state has an intrinsic error 
  of $E=5.5\times 10^{-5}$ at $R=5.~\mu\rm m$ using $\Omega/2\pi \simeq 30~\rm MHz$.

\subsection{Technical errors}

The main technical errors that affect the $C_Z$ operation are 
spontaneous emission during Rydberg excitation, magnetic noise and Doppler effects. The spontaneous emission probability can always be made sufficiently small provided sufficient laser power is available for the Rydberg excitation step, although this may be a technical challenge. For example let us consider the feasibility of achieving  a 30 MHz excitation frequency to the $150s_{1/2}$ state while maintaining  $< 10^{-3}$ probability of spontaneous emission in a Rydberg $\pi$ pulse. In a two-photon excitation scheme spontaneous emission 
is minimized by setting the one-photon Rabi frequencies equal, in which case the probability of emission from the intermediate $p$ level during a $\pi$ excitation pulse is related to the two-photon Rabi frequency by \cite{Saffman2010}
$$
\Omega=\frac{P_{\rm se}}{\pi}\frac{|\Omega_1|^2}{\gamma_p},
$$
with $\Omega_1$ the one-photon Rabi frequency and $\gamma_p$ the radiative decay rate of the $p$ level. 
 
Some sample numbers for excitation through $5p_{1/2}$ are 
$20~\mu\rm W$ of 795 nm light,  $10 ~\rm W$ of 474 nm light focused to beam waists 
of $1~\mu\rm m$ using  a detuning of 37 GHz, which gives 
$\Omega/2\pi=30 ~\rm MHz$ and $P_{\rm se}=3\times 10^{-4}$.
Generating such high single frequency power at 474 nm would probably require using a pulsed amplifier that is then frequency doubled, as in\cite{Mamaev2003}.  This would be technically challenging in comparison to the cw lasers used in current Rydberg Rabi oscillation experiments\cite{Johnson2008}. An alternative is to perform excitation through the second resonance level, $5s_{1/2}\rightarrow 6p_{1/2}\rightarrow ns_{1/2}$ using 422 and 1004 nm lasers, as has been demonstrated in \cite{Viteau2010}.  The smaller value of $\gamma_{6p_{1/2}}=1/125~\rm ns$, and larger matrix element to the Rydberg level leads to excitation of $150s_{1/2}$ with 
$\Omega/2\pi=30 ~\rm MHz$ and $P_{\rm se}=1\times 10^{-4}$ using $5~\mu\rm W$ of 422 nm light and 2 W of 1004 nm light. The high power requirement is thereby moved to the infrared which is technically more convenient.

Doppler broadening due to atomic motion leads to imperfect Rydberg excitation. This is a small effect provided $\Delta_D\ll \Omega$ where
$\Delta_D={\bf k}_{2\nu}\cdot{\bf v}$ is the Doppler shift, $\bf v$ is the atomic velocity and ${\bf k}_{2\nu}={\bf k}_1-{\bf k}_2$ is the two-photon excitation wavenumber. For example at $\Omega/2\pi=30~\rm MHz$ and $T=100~\mu\rm K$ atoms the probability of an error in exciting a Rydberg  atom is under $10^{-5}.$

\begin{figure}[!t]
\vspace{-1.1cm}
\begin{center}
\includegraphics[width=9cm]{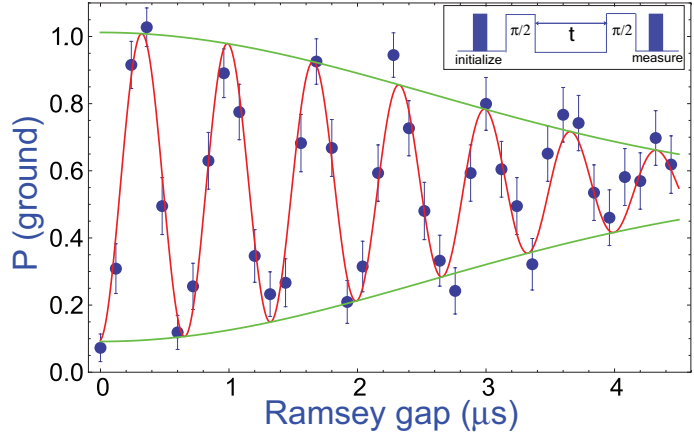}
\end{center}
\vspace{-.7cm}
\caption{\small \label{fig.Ramsey}Ramsey experiment between ground and Rydberg states. A fit to the envelope gives a decay time of $T_2=3.6~\mu\rm s$. }
\end{figure}
 
A more serious technical error arises from dephasing of the Rydberg state relative to the ground state during gate operation. We have observed this dephasing by performing a Ramsey interference experiment between the ground state $\ket{g}=\ket{5s_{1/2},f=2,m_f=0}$  and the 
Rydberg state $\ket{r}=\ket{97d_{5/2},m_j=5/2}$. The results are shown in Fig. \ref{fig.Ramsey}. A short analysis shows that the envelope of the Ramsey oscillations is given by $\langle e^{\imath\varphi_{\rm st}}\rangle$ with $\varphi_{\rm st}$ any stochastic phase that is accumulated during the Ramsey gap time $t.$

The stochastic phase is due to two main sources, magnetic field fluctuations and motional effects. Modeling magnetic field fluctuations as a random Gaussian process with standard deviation $\sigma$ we find 
$$
\langle e^{\imath\varphi_{\rm st}}\rangle = e^{-t^2/T_{2,B}^2}
$$
where 
$$
T_{2,B}= \frac{2^{3/2}\pi\hbar}{|g_R m_{jR}-g_g m_{fg}| \mu_B \sigma} 
$$
where $|g_R m_{jR}-g_g m_{fg}| \mu_B$ is the difference in Zeeman shift per unit field strength between Rydberg fine structure and ground hyperfine states with $\mu_B$ a Bohr magneton.  This  definition of $T_2$ corresponds to the $1/e$ decay time of the Ramsey fringes. 

Doppler broadening, in addition to imperfect Rydberg excitation,  leads to 
motional decoherence of the Rydberg state according to 
$$
\langle e^{\imath\varphi_{\rm st}}\rangle = e^{-t^2/T_{2,D}^2}
$$
where 
$$
T_{2,D}=\left(\frac{2m}{k_BT} \right)^{1/2}\frac{1}{k_{2\nu}}
$$
with $m$ the atomic mass, $T$ the temperature, and $k_B$ Boltzmann's constant. A physical picture of this decoherence mechanism is that motion of the Rydberg excited atom causes it to see a different phase of the two-photon field 
which gets imprinted onto the wavefunction upon deexcitation.

In the  presence of both magnetic and Doppler dephasing the effective $T_2$ is 
$$
T_2=  \frac{T_{2,B} T_{2,D}}{(T_{2,B}^2+T_{2,D}^2)^{1/2}} .
$$
The  decoherence due to magnetic and Doppler effects can then be extracted from Fig. \ref{fig.Ramsey} if we know the magnetic field fluctuations $\sigma$. The field fluctuations were estimated  to be
$\sigma\sim 2.5\times 10^{-6}~\rm T$
  by measuring $T_2$ of the hyperfine qubit at two different bias magnetic field strengths.
Figure \ref{fig.Ramsey} then implies that   $T\sim 60~\mu\rm K, $
 $T_{2,B}=13~\mu\rm s$ and $T_{2,D}=3.7~\mu\rm s$. We see that the 
Rydberg decoherence is dominated by Doppler dephasing in these experiments. 

\begin{figure}[!t]
\vspace{-1.1cm}
\begin{center} 
\includegraphics[width=13cm]{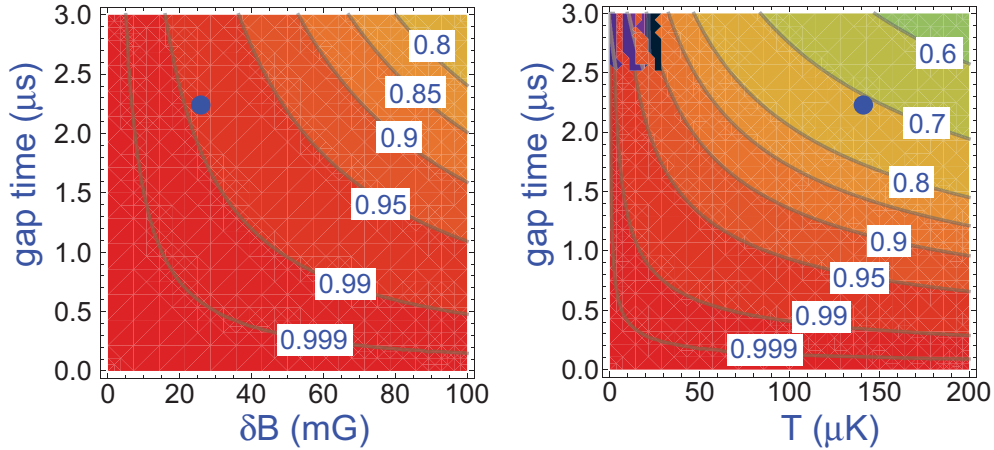}
\end{center}
\vspace{-.7cm}
\caption{\small \label{fig.fidelity}Bell state fidelity limits set by magnetic and Doppler dephasing. Calculations are for $^{87}$Rb with $
k_{2\nu}=k_{480~\rm nm}-k_{780~\rm nm}.$ The blue dots correspond to the experiments of \cite{Zhang2010}. }
\end{figure}

Dephasing of  the Ramsey interference signal also limits the 
fidelity of the entanglement that is created using Rydberg interactions. Averaging over the phase $\varphi$ in Eq. (\ref{eq.Bell}) leads  to
\be
F=\frac{1}{2}\left(1+\langle e^{\imath\varphi_{\rm st}}\rangle \right).
\label{eq.Flimit}
\ee 
This limit  was first pointed out  in \cite{Wilk2010}. Figure \ref{fig.fidelity} shows the fidelity limits set by magnetic and Doppler dephasing. These limits are in agreement with the atom loss corrected fidelity of $F=0.71$ reported in \cite{Zhang2010}. 

 \subsection{Discussion} 

The most accurate quantum gate experiments to date are those of the Innsbruck ion trap group who have reported two-qubit entanglement fidelities better than 99\%\cite{Benhelm2008b}. Can neutral atom Rydberg gates hope to reach similar fidelities?

The discussion in the previous sections has identified  the leading error sources as being due to
intrinsic gate errors, Doppler excitation error, spontaneous emission, magnetic dephasing, and 
Doppler dephasing. Assuming excitation of $150s_{1/2}$ states with $\Omega/2\pi= 30 ~\rm MHz$   the intrinsic gate error,  Doppler excitation error, and spontaneous emission error can all be  under $10^{-4}$ as discussed above. The minimum  gap time during which the control atom is Rydberg excited is $2\pi/\Omega = .03~\mu\rm s$. Looking at Fig. \ref{fig.fidelity} we see that such short gap times would lead to dephasing errors well below $10^{-3}$ even with our current values of magnetic noise and atomic temperature. 

The remaining technical challenges concern accurate timing with 30 MHz Rabi frequencies as well as sufficient laser power as discussed above. A $10^{-3}$  error in the population of a target state on a $\pi$ pulse at 30 MHz requires timing accuracy at the level of about 1 ns, which is feasible with available technology. An interim conclusion is that gate errors at the $10^{-3}$ level are possible, but will require  careful attention to technical issues.

\section{Outlook}
\label{sec.outlook}

The utility of neutral atom qubits for quantum information processing depends on more than high gate fidelities. For circuit model quantum computing it is also necessary to develop architectures which support site selective gates, initialization, and measurements for a many-qubit array. Although techniques for loading millions of atoms into optical lattices are well developed there are specific challenges associated with Rydberg blockade quantum computing. 

One of the foremost issues is that red detuned dipole traps can store ground state atoms with low decoherence but they do not trap Rydberg states. The polarizability of a highly excited Rydberg atom is essentially that of a free electron, and is negative. Thus a red detuned trap presents repulsive mechanical potential for the Rydberg atom. Furthermore the trapping light rapidly photoionizes the Rydberg atom\cite{Saffman2005a}. This problem is handled in current experiments\cite{Wilk2010,Isenhower2010,Zhang2010} by turning off the trapping light for a few $\mu\rm s$ during Rydberg excitation. Although this leads to a small amount of mechanical heating and decoherence due to entanglement between the qubit state and the center of mass motion, this has not been a significant issue until now. 

In a multi-qubit array turning off the traps whenever a Rydberg gate is performed could lead to a significant heating and decoherence problem. One solution would be to work with an optical system where each trap could be turned on and off individually on $\mu\rm s$ time scales. This is not compatible  with multi-trap architectures based on
multi-beam optical lattices\cite{Nelson2007}, 
or fixed diffraction gratings\cite{Knoernschild2010}, but can be achieved at few ms time scales using  spatial light modulator based approaches\cite{Bergamini2004,Kruse2010}. 

A solution which negates the need to turn the traps off is to work at a wavelength where the trap is attractive for both ground state and Rydberg atoms. This can be achieved in a blue detuned optical lattice\cite{Nelson2007} or a blue detuned bottle beam trap\cite{Isenhower2009}. Bottle beam traps specifically designed for ``magic" trapping of ground and Rydberg atoms are currently under development at UW Madison. It is also necessary to consider Rydberg state photoionization rates due to the trapping light. In a blue detuned trap the nucleus is in a region of  low light intensity, and since the photoionization matrix element is dominated by the  region near the core, the rate tends to be strongly suppressed\cite{Robicheaux2010}.

All of these issues are challenging, but also susceptible to solution. We believe that the combination of potentially high gate fidelity, and suitable architectures for many-qubit implementations, bode well for the future development of Rydberg blockade based quantum computing.

This work was supported by the NSF under grant PHY-1005550, IARPA through ARO contract W911NF-10-1-0347, and DARPA under award FA8750-09-1-0230. 

\section*{References}

%
%


\begin{thebibliography}{10}
\expandafter\ifx\csname url\endcsname\relax
  \def\url#1{{\tt #1}}\fi
\expandafter\ifx\csname urlprefix\endcsname\relax\def\urlprefix{URL }\fi
\providecommand{\eprint}[2][]{\url{#2}}

\bibitem{Treutlein2004}
Treutlein P, Hommelhoff P, Steinmetz T, H\"ansch T~W and Reichel J 2004 {\em
  Phys. Rev. Lett.\/} {\bf 92} 203005

\bibitem{Brennen1999}
Brennen G~K, Caves C~M, Jessen P~S and Deutsch I~H 1999 {\em Phys. Rev.
  Lett.\/} {\bf 82} 1060

\bibitem{Jaksch1999}
Jaksch D, Briegel H~J, Cirac J~I, Gardiner C~W and Zoller P 1999 {\em Phys.
  Rev. Lett.\/} {\bf 82} 1975

\bibitem{Pellizzari1995}
Pellizzari T, Gardiner S~A, Cirac J~I and Zoller P 1995 {\em Phys. Rev.
  Lett.\/} {\bf 75} 3788

\bibitem{You2000}
You L and Chapman M~S 2000 {\em Phys. Rev. A\/} {\bf 62} 052302

\bibitem{Lukin2000}
Lukin M~D and Hemmer P~R 2000 {\em Phys. Rev. Lett.\/} {\bf 84} 2818--2821

\bibitem{Mompart2003}
Mompart J, Eckert K, Ertmer W, Birkl G and Lewenstein M 2003 {\em Phys. Rev.
  Lett.\/} {\bf 90} 147901

\bibitem{Mandel2003}
Mandel O, Greiner M, Widera A, Rom T, H\"ansch T~W and Bloch I 2003 {\em Nature
  (London)\/} {\bf 425} 937--940

\bibitem{Anderlini2007}
Anderlini M, Lee P~J, Brown B~L, Sebby-Strabley J, Phillips W~D and Porto J~V
  2007 {\em Nature (London)\/} {\bf 448} 452--456

\bibitem{Jaksch2000}
Jaksch D, Cirac J~I, Zoller P, Rolston S~L, C\^ot\'e R and Lukin M~D 2000 {\em
  Phys. Rev. Lett.\/} {\bf 85} 2208--2211

\bibitem{Nielsen2000}
Nielsen M~A and Chuang I~L 2000 {\em Quantum computation and quantum
  information\/} (Cambridge University Press, Cambridge)

\bibitem{Hagley1997}
Hagley E, Ma\^itre X, Nogues G, Wunderlich C, Brune M, Raimond J~M and Haroche
  S 1997 {\em Phys. Rev. Lett.\/} {\bf 79} 1

\bibitem{Saffman2010}
Saffman M, Walker T~G and M\o{}lmer K 2010 {\em Rev. Mod. Phys.\/} {\bf 82}
  2313

\bibitem{Saffman2005a}
Saffman M and Walker T~G 2005 {\em Phys. Rev. A\/} {\bf 72} 022347

\bibitem{Singer2004}
Singer K, Reetz-Lamour M, Amthor T, Marcassa L~G and Weidem\"uller M 2004 {\em
  Phys. Rev. Lett.\/} {\bf 93} 163001

\bibitem{Tong2004}
Tong D, Farooqi S~M, Stanojevic J, Krishnan S, Zhang Y~P, C\^ot\'e R, Eyler E~E
  and Gould P~L 2004 {\em Phys. Rev. Lett.\/} {\bf 93} 063001

\bibitem{Urban2009}
Urban E, Johnson T~A, Henage T, Isenhower L, Yavuz D~D, Walker T~G and Saffman
  M 2009 {\em Nature Phys.\/} {\bf 5} 110

\bibitem{Gaetan2009}
Ga\"etan A, Miroshnychenko Y, Wilk T, Chotia A, Viteau M, Comparat D, Pillet P,
  Browaeys A and Grangier P 2009 {\em Nature Phys.\/} {\bf 5} 115

\bibitem{Lukin2001}
Lukin M~D, Fleischhauer M, Cote R, Duan L~M, Jaksch D, Cirac J~I and Zoller P
  2001 {\em Phys. Rev. Lett.\/} {\bf 87} 037901

\bibitem{Wilk2010}
Wilk T, Ga\"etan A, Evellin C, Wolters J, Miroshnychenko Y, Grangier P and
  Browaeys A 2010 {\em Phys. Rev. Lett.\/} {\bf 104} 010502

\bibitem{Gaetan2010}
Ga\"etan A, Evellin C, Wolters J, Grangier P, Wilk T and Browaeys A 2010 {\em
  New J. Phys.\/} {\bf 12} 065040

\bibitem{Isenhower2010}
Isenhower L, Urban E, Zhang X~L, Gill A~T, Henage T, Johnson T~A, Walker T~G
  and Saffman M 2010 {\em Phys. Rev. Lett.\/} {\bf 104} 010503

\bibitem{Zhang2010}
Zhang X~L, Isenhower L, Gill A~T, Walker T~G and Saffman M 2010 {\em Phys. Rev.
  A\/} {\bf 82} 030306(R)

\bibitem{Walker2008}
Walker T~G and Saffman M 2008 {\em Phys. Rev. A\/} {\bf 77} 032723

\bibitem{Mamaev2003}
Mamaev A~V, Lodahl P and Saffman M 2003 {\em Opt. Lett.\/} {\bf 28} 31

\bibitem{Johnson2008}
Johnson T~A, Urban E, Henage T, Isenhower L, Yavuz D~D, Walker T~G and Saffman
  M 2008 {\em Phys. Rev. Lett.\/} {\bf 100} 113003

\bibitem{Viteau2010}
Viteau M, Radogostowicz J, Chotia A, Bason M~G, Malossi N, Fuso F, Ciampini D,
  Morsch O, Ryabtsev I~I and Arimondo E 2010 {\em J. of Phys. B: At., Mol. and
  Opt. Phys.\/} {\bf 43} 155301

\bibitem{Benhelm2008b}
Benhelm J, Kirchmair G, Roos C~F and Blatt R 2008 {\em Nat. Phys.\/} {\bf 4}
  463

\bibitem{Nelson2007}
Nelson K~D, Li X and Weiss D~S 2007 {\em Nat. Phys.\/} {\bf 3} 556--560

\bibitem{Knoernschild2010}
Knoernschild C, Zhang X~L, Isenhower L, Gill A~T, Lu F~P, Saffman M and Kim J
  2010 {\em Appl. Phys. Lett.\/} {\bf 97} 134101

\bibitem{Bergamini2004}
Bergamini S, Darqui\'{e} B, Jones M, Jacubowiez L, Browaeys A and Grangier P
  2004 {\em J. Opt. Soc. Am. B\/} {\bf 21} 1889--1894

\bibitem{Kruse2010}
Kruse J, Gierl C, Schlosser M and Birkl G 2010 {\em Phys. Rev. A\/} {\bf 81}
  060308

\bibitem{Isenhower2009}
Isenhower L, Williams W, Dally A and Saffman M 2009 {\em Opt. Lett.\/} {\bf 34}
  1159

\bibitem{Robicheaux2010}
F. Robicheaux, private communication.

\end{thebibliography}

\providecommand{\newblock}{}


\end{document}